\documentclass[prb ,aps,twocolumn,groupaddress]{revtex4-2}
\usepackage{titlesec}
\titleformat*{\section}{\bf\fontsize{12}{2}\selectfont}
\usepackage[latin1]{inputenc}
\usepackage{amsmath}
\usepackage{amsfonts}
\usepackage{bm}
\usepackage{amssymb}
\usepackage{graphicx,bm}
\usepackage{mathrsfs}
\usepackage{hyperref}
\usepackage{esint}
\usepackage[caption=false]{subfig}

\newcommand{\bpm}{\begin{pmatrix}}
\newcommand{\lp}{\left(}

\newcommand{\lb}{\left[}

\newcommand{\rb}{\right]}

\newcommand{\rp}{\right)}
\newcommand{\e}{\epsilon}

\newcommand{\epm}{\end{pmatrix}}

\newcommand{\ba}{\begin{eqnarray}}
\newcommand{\ban}{\begin{eqnarray}\nn}
\newcommand{\ea}{\end{eqnarray}}
\newcommand{\nn}{\nonumber}
\newcommand{\im}{\text{Im }}
\newcommand{\re}{\text{Re }}
\begin{document}
\author{Christian Boyd, Luke Yeo, and Philip W. Phillips}
\affiliation{Department of Physics and Institute for Condensed Matter Theory,
	University of Illinois 1110 W. Green Street, Urbana, IL 61801, U.S.A}
\title{Probing the bulk plasmon continuum of layered materials through electron energy loss spectroscopy in a reflection geometry}

\begin{abstract}
A periodic arrangement of 2D conducting planes is known to host a (bulk) plasmon dispersion that interpolates between the typical, gapped behavior of 3D metals and a gapless, acoustic regime as a function of the out-of-plane wavevector.  The semi-infinite system --- the configuration relevant to Electron Energy Loss Spectroscopy (EELS) in a reflection geometry, as in High Resolution EELS (HREELS) --- is known to host a surface plasmon that ceases to propagate below a cutoff wavevector.  As the f-sum rule requires a finite response whether or not there exist sharp excitations, we demonstrate that what remains in the surface loss function --- the material response probed by HREELS --- is the contribution from the (bulk) plasmon of the infinite system.  In particular, we provide a one-to-one mapping between the plasmon continuum and the spectral weight in the surface loss function.  In light of this result, we suggest that HREELS be considered a long wavelength probe of the plasmon continuum in layered materials.
\end{abstract}
\maketitle

\section*{Introduction}

\begin{figure}[h!]
	\centering
	\subfloat[\label{sfig:InPhase}]{
		\includegraphics[width=0.49\linewidth]{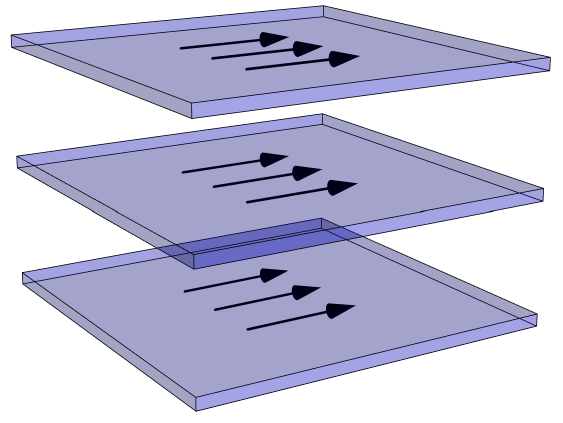}
	}
	\subfloat[\label{sfig:OutOfPhase}]{
		\includegraphics[width=0.49\linewidth]{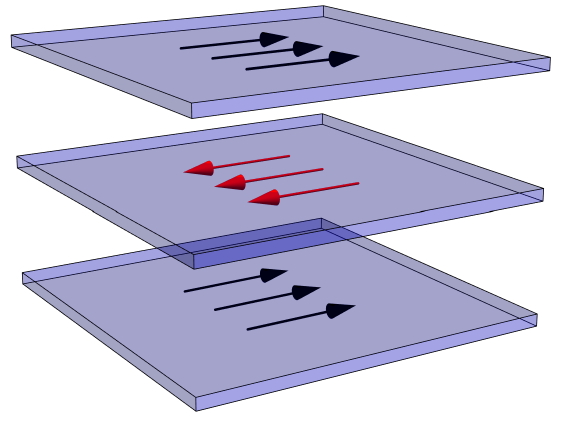}
	}
	
	\caption{A visual representation of the {\bf(a)} in-phase and {\bf(b)} out-of-phase charge oscillation within a periodic system of conducting planes.}
	\label{figOscillation}
\end{figure}

\begin{figure}[h!]
	\centering
	\includegraphics[width=\linewidth]{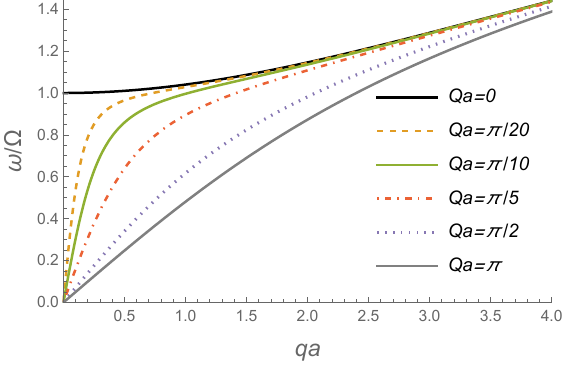}
	\caption{Plotted above is the dispersion of the plasma frequency $\omega_p$ \eqref{bulk dispersion} for a periodic system of 2D conducting planes.  The frequency scale is in units of the optical plasma frequency $\Omega$ \eqref{Omega def}.  The in-plane wavevector $q$ and out-of-plane wavevector $Q$ are given in units of the inter-plane separation $a$.}
	\label{figBulk}
\end{figure}

In the mid 1970s, A. L. Fetter \cite{Fetter1974} applied a hydrodynamic analysis to a periodic system of conducting planes,
each hosting a 2D electron gas, and obtained a rather
unique plasmon continuum. The plasmon -- the long
wavelength oscillation of the many-electron charge density -- radically changes in character depending on
whether adjacent conducting planes of the system are
oscillating in-phase or out-of-phase. While the in-phase
oscillation (see Fig. \ref{sfig:InPhase}) corresponds to a gapped (optical) mode reminiscent of plasmons in simple 3D metals,
the out-of-phase oscillation (see Fig. \ref{sfig:OutOfPhase}) corresponds
to a gapless (acoustic) mode that disperses linearly with in-plane
wavevector. Between the
extremes of in-phase and out-of-phase oscillation lies an
acoustic-to-optical plasmon continuum (see Fig. \ref{figBulk}).

Due to the severe conduction anisotropy in the high-$T_C$ cuprate superconductors \cite{Ando2002,Komiya2002,Watanabe1997,Boebinger1996}, doped cuprates are often modeled as a system of copper oxide planes --- i.e., the model studied by Fetter.  Significant experimental effort has recently been invested toward demonstrating that doped cuprates exemplify the plasmon continuum scenario using Resonant Inelastic X-ray Scattering (RIXS) on both electron-doped cuprates, such as La$_{2-x}$Ce$_x$CuO$_4$ and Nd$_{2-x}$Ce$_x$CuO$_4$ \cite{Hepting2018,Lin2020}, as well as the hole-doped compounds La$_{2-x}$Sr$_x$CuO$_4$ (LSCO) and Bi$_{2}$Sr$_{1.6}$La$_{0.4}$CuO$_{6+\delta}$ (Bi-2201) \cite{Abhishek2020,Singh2020}.  Shared among these RIXS studies is a peak in the scattered intensity whose (planar) wavevector dispersion changes appreciably as the out-of-plane wavevector is tuned toward an out-of-phase oscillation between adjacent copper oxide layers.  By identifying this peak with the plasmon, RIXS data provides strong evidence of a nearly-acoustic mode in doped cuprates coincident with the out-of-phase, acoustic plasmon dispersion of the Fetter model.  This analogy to the Fetter model can be completed by tracking the evolution toward the optical plasmon dispersion when the out-of-plane wavevector corresponds to in-phase oscillation; however, the scattering geometry and suppression of charge excitations at large out-of-plane wavevector in the RIXS cross section make this connection difficult to establish concretely \cite{Abhishek2020comment}.

Rather than following the (3D) wavevector dispersion of charge density excitations, we instead consider an experimental probe that lacks translation symmetry along a chosen axis: Electron Energy Loss Spectroscopy (EELS) in a {\it reflection} geometry; i.e., High Resolution EELS (HREELS).  When the reflection surface is perpendicular to the out-of-plane (layering) direction, the explicitly broken translation symmetry renders HREELS, in principle, susceptible to {\it all} out-of-plane wavevector contributions-- a point previously made in \cite{Schulte1999,Schulte2002}.  Our main result is that EELS in a reflection geometry (HREELS) does, in fact, probe the \textit{bulk} plasmon continuum of the Fetter model in the long wavelength limit.  As a consequence, layered materials should generically be expected to host a broad surface response at long wavelengths due to their underlying bulk plasmon continuum.

While the doped cuprates provide motivation for this investigation, these compounds are notoriously complex.  Because of its relative simplicity and close correspondence to the plasmon behavior seen in RIXS, we will instead focus on the Fetter model of layered conducting planes.  Though the aforementioned studies \cite{Hepting2018,Lin2020,Abhishek2020,Singh2020} contain more involved computational analyses to fit their experimental curves, the qualitative plasmon behavior (or treatment of the Coulomb interaction) often mirrors the Fetter result \cite{Fetter1974}.  One could argue that the Fetter model might represent a pathological limit since it neglects inter-plane conduction; however, the inclusion of inter-layer electron hopping results in a plasmon dispersion that continuously evolves from the Fetter result to a more general anisotropic dispersion \cite{Grecu1973}.  This is to say that our results apply inasmuch as the Fetter model is capable of describing any particular (strongly) anisotropic system.

\section{The semi-infinite Fetter model}
\label{section 1}
\begin{figure}[h!]
	\centering
	\includegraphics[width=\linewidth]{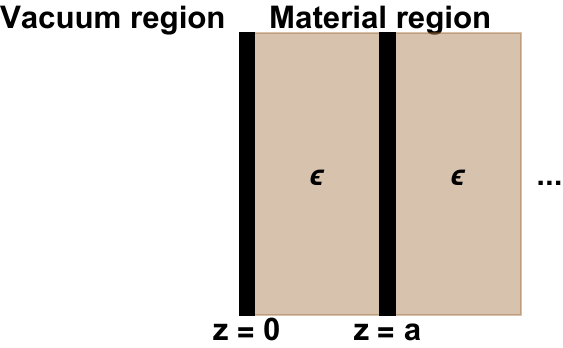}
	\caption{The semi-infinite model of periodic conducting planes --- the semi-infinite Fetter model --- considered in the text.  The thick black lines denote 2D conducting planes, the dielectric constant within the insulating layers is denoted by $\e$, and the layer periodicity is given by $a$.}
	\label{figLattice}
\end{figure}

In our implementation of the Fetter model, we consider a system of conducting planes separated by (insulating) dielectric layers --- i.e., a semi-infinite, single-layer superlattice (see Fig. \ref{figLattice}).  The conducting planes are characterized by a long wavelength, 2D Drude conductivity $\sigma$ at frequency $\omega$ through the standard relation
\ba
\label{Drude conductivity}
\sigma(\omega) = \frac{ine^2}{m\omega\lp1+i/\omega\tau\rp}\,\,\,,
\ea
where $n$ is the 2D (planar) electron density, $e$ the electron charge, $m$ the electron mass, and $\tau$ corresponds to a relaxation time associated with electron scattering.  The layer periodicity of the 2D conducting planes (or the inter-layer distance) is denoted by $a$ and we model the insulating, dielectric regions through a dielectric constant $\e>1$.

The excitations of the infinite Fetter model --- the periodic system of conducting planes without boundary --- are well studied.  In the limit of negligible conduction dissipation ($\omega_p\tau\gg1$), the bulk plasmon dispersion with the in-plane wavevector $q$ and out-of-plane wavevector $Q$ is given by \cite{Fetter1974,Cottam1986,Cottam2004,Jain1985_2,Apostol1975,Quinn1983}
\ba
\label{bulk dispersion}
\omega_p(q,Q) = \Omega\sqrt{\frac{qa\sinh qa}{2\cosh qa-2\cos Qa}}
\,\,\,,
\ea
where $\Omega$ is the optical plasma frequency defined as
\ba
\label{Omega def}
\Omega^2:=\frac{ne^2}{\e_0\e ma}\,\,\,.
\ea
The $\omega_p$ dispersion \eqref{bulk dispersion} is plotted in Fig. \ref{figBulk}.  

\begin{figure}[h!]
	\centering
	\includegraphics[width=\linewidth]{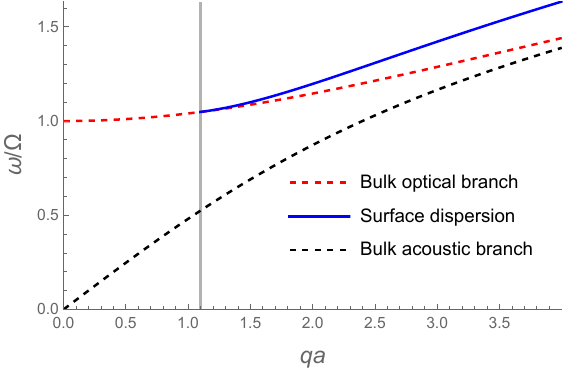}
	\caption{Plotted above is the dispersion of the surface plasma frequency $\omega_{sp}$ \eqref{surface dispersion} in the semi-infinite Fetter model (solid line) for the value $\e=2$.  The bulk optical and acoustic branches (dashed lines) are obtained from the plasma frequency $\omega_p(q,Q)$ \eqref{bulk dispersion} at $Qa=0$ and $Qa=\pi$, respectively.  The vertical bar denotes the cutoff wavevector $q^*a$, determined by $\e$ through \eqref{cutoff q}, below which the surface plasmon ceases to propagate.}
	\label{figSurface}
\end{figure}

In the semi-infinite geometry (i.e., as in Fig. \ref{figLattice}), bulk plasmons are no longer self-sustained resonances of the system due to the lack of translation symmetry in the out-of-plane direction.  Instead, {\it surface} plasmons --- planar charge density oscillations that decay away from the vacuum-material interface --- are the long-lived excitations of the superlattice.  When corrections due to the finite relaxation time $\tau$ are negligible, the surface modes follow the dispersion relation \cite{Quinn1983,Jain1985_2,Cottam2004}
\ba
\label{surface dispersion}
\omega_{sp}(q) = \Omega\sqrt{\frac{qa \e \lb
\e\cosh qa -\sinh qa
\rb}{\lp\e^2-1\rp \sinh qa}}
\,\,\,,
\ea
whose functional form depends on the dielectric mismatch at the vacuum-material interface.  The $\omega_{sp}$ dispersion \eqref{surface dispersion} is plotted in Fig. \ref{figSurface}.  Curiously, the surface plasmon {\it ceases to exist} for sufficiently long wavelengths and is never a resonance of the system in the absence of a dielectric background \cite{Mahan2012}.  The cutoff wavevector $q^*$,
\ba
\label{cutoff q}
q^*a:= \ln\lp\frac{\e+1}{\e-1}\rp
\,\,\,,
\ea
marks the lower bound for surface plasmon propagation at a given $\e>1$ \cite{Quinn1983,Jain1985_2,Cottam1986}.

\section{Dielectric analysis}
The lack of long-lived charge density excitations in the $q<q^*$ regime of the semi-infinite Fetter model poses an interesting question for a High Resolution Electron Energy Loss Spectroscopy (HREELS) experiment: \textit{What does HREELS measure at wavevectors below $q^*$?}  Independent of sharp excitations, the spectral weight in the (dissipative) density response for $q<q^*$ is nevertheless constrained by the f-sum rule in a reflection geometry \cite{Gumhalter1975,Griffin1976,Eguiluz1981,Gumhalter1984,Liebsch1997,Uchoa2016}.  Put simply, HREELS has to measure {\it something} associated with charge density excitations at long wavelengths --- and a fixed amount of it.  It has previously been found that, in addition to the surface plasmon, there is a finite contribution from the bulk plasmon \cite{Bergara1999,Pitarke2001}; however, the bulk correction typically constitutes a weak or visually-imperceptible shoulder in the surface loss function above the (bulk) plasma frequency.  In order to quantify the redistribution of spectral weight in the $q<q^*$ regime of the semi-infinite Fetter model, we study the surface loss function
\ba
\nn
g(q,\omega)&:=&-\frac{e^2}{2\e_0|q|}\int_0^\infty dz dz' \chi''(q,\omega;z,z') e^{-|q|z} e^{-|q|z'}
\\ \label{surface loss function} &\,&
\ea
since this {\it dimensionless} quantity captures the material response contribution to the HREELS cross section \cite{Ritchie1959,Mills1975,Liebsch1997,MEELS}.  In \eqref{surface loss function}, $\chi''$ denotes the imaginary part of the density response function $\chi$ which has been Fourier transformed along the planar and temporal directions ---  i.e., we assume planar translation invariance --- and the material region is chosen as the half-space $z>0$.

As our implementation of the semi-infinite Fetter model (see Fig. \ref{figLattice}) is made up of alternating conducting and dielectric layers, we calculate the surface loss function in \eqref{surface loss function} through a dielectric analysis of the system.  In the dielectric theory of reflection EELS \cite{Mills1972,Mills1975,Mills1984,Lucas1972} (and the vacuum scattering contributions in \cite{Ritchie1957,Ritchie1959}), one characterizes the system through local dielectric response (in the out-of-plane, or layering, direction) and derives the long-ranged effect of the Coulomb interaction by enforcing electromagnetic boundary conditions across each interface.  Superlattice boundary conditions are often written in the full electrodynamic formalism in terms of the electric and magnetic field vectors \cite{Cottam1986,Cottam1989,Cottam2004}.  Instead, we take the non-retarded limit $q^2\gg\omega^2/c^2$ from the outset and work in a theory of potentials $\phi,\phi_D$ for the electric field $E=-\nabla\phi$ and the electric displacement $D=-\e_0\nabla \phi_D$, which are themselves related in the dielectric regions through $D=\e_0\e E$.  At each interface, we have the standard boundary conditions that the planar components of $E$ are continuous and, across a conducting plane, that the discontinuity in the $D_z$ is given by the planar charge density.  By use of the continuity equation and recognizing that the finite ($q,\omega$) components of the planar charge density are entirely due to the induced response, the discontinuity in $D_z$ can be related to the planar components of $E$ and the conductivity $\sigma$.  Lastly, the infinite system of boundary conditions can be closed by requiring decaying (bounded) behavior for the $\phi,\phi_D$ within the material (e.g., see \cite{Quinn1983,Cottam2004}).

The dielectric analysis provides calculational utility since the surface loss function \eqref{surface loss function} is encoded in the vacuum solution for the electric potential when a driving field is applied \cite{Liebsch1997,Persson1985,Mills1984,Nazarov1994}.  To be concrete, consider a semi-infinite system whose boundary is {\it defined} by a barrier to its charge density at $z=0$, as in our Fetter model of Fig. \ref{figLattice}, and is embedded within a dielectric medium characterized by $\e$.  If a perturbing potential $\phi_{ext}$ is sourced by a charge density localized entirely within the vacuum region (implementing the negligible material penetration assumption used to construct the HREELS cross section \cite{Mills1972,Mills1975,MEELS}), then the material responds to a field of the form
\ba
\label{phi ext}
\phi_{ext}(q,\omega,z>0) = C_{ext}(q,\omega) e^{-|q|z}
\ea
by inducing the charge density
\ba
 \nn
\rho_{ind}(q,\omega,z) &=& e^2C_{ext}(q,\omega)\int_{z'>0} dz' \chi(q,\omega;z,z') e^{-|q|z'}
\\ \label{induced charge density} &\,&
\ea
via the definition of the density response function $\chi$.  Outside of the material (in the vacuum region $z<0$), the induced charge density \eqref{induced charge density} sources an induced potential
\ba
\label{induced potential}
&\,&\phi_{ind}(q,\omega,z<0) = \frac{e^2}{2\e_0|q|}
\lp\frac{2}{1+\e}\rp
\\ \nn &\,&
\times C_{ext}(q,\omega) e^{+|q|z}\int_0^\infty dz' dz'' \chi(q,\omega;z',z'') e^{-|q|z'}e^{-|q|z''}
\,\,\,.
\ea
In \eqref{induced potential}, $\phi_{ind}$ depends on $\e$ through the boundary conditions across $z=0$.  Already, the induced potential in vacuum \eqref{induced potential} is probing the density response function in the same way as the surface loss function \eqref{surface loss function}.  

In order to extract the surface loss function from the induced potential \eqref{induced potential}, we require an explicit form of the external field coefficient $C_{ext}$ in \eqref{phi ext}.  A convenient external field --- $\phi_{ext}$ of \eqref{phi ext} --- to consider is the classical dielectric description of EELS in a reflection geometry \cite{Lucas1971,Mills1984,Lucas1985} wherein an incoming electron is modeled by a trajectory that undergoes elastic reflection off the material surface ($z=0$) at the time $t=0$.  The potential $\phi_{ext}$ sourced by this classical electron trajectory is
\ba
\nn
\phi_{ext}(q,\omega,z<0) &=&
A(q,\omega)\cos\lb
\lp\omega-q\cdot v_\parallel\rp\lp z/v_z\rp
\rb
\\ \label{phi ext vac} &\,&
-\lp\frac{\e}{1+\e}\rp A(q,\omega)e^{+|q|z}
\ea
in the vacuum region and
\ba
\label{phi ext mat}
\phi_{ext}(q,\omega,z>0) &=&
\lp\frac{1}{1+\e}\rp A(q,\omega)e^{-|q|z}
\ea
in the material region.  In \eqref{phi ext vac} and \eqref{phi ext mat}, dependence on $\e$ is through the boundary conditions across $z=0$, the coefficient $A$ is defined by
\ba
\label{A def}
A(q,\omega):=-\frac{2e v_z}{\e_0\lb
q^2 v_z^2+\lp\omega-q\cdot v_\parallel\rp^2
\rb}\,\,\,,
\ea
and the reflection trajectory of the classical electron is parameterized by its velocity components $v_{\parallel}$ and $v_z$, which are, respectively, parallel and perpendicular to the ($z=0$) surface.  In light of \eqref{phi ext mat}, we can immediately write down the material response (in vacuum) to this particular $\phi_{ext}$ using \eqref{induced potential} through the replacement $C_{ext}\to A/(1+\e)$ as
\ba
\label{phi vac response}
&\,&
\phi_{ind}(q,\omega,z<0)
=\frac{e^2}{2\e_0|q|} \frac{2}{(1+\e)^2} A(q,\omega) e^{+|q|z}
\\ \nn &\,&
\times\int_0^\infty dz' dz'' \chi(q,\omega;z',z'') e^{-|q|z'}e^{-|q|z''}
\,\,\,.
\ea

The vacuum response of the material \eqref{phi vac response} provides a recipe to calculate the surface loss function within a dielectric model.  Since $A$ \eqref{A def} is real (and, as a result, so is $\phi_{ext}$ \eqref{phi ext vac} in vacuum), the imaginary part of the total potential $\phi_{tot}:=\phi_{ind}+\phi_{ext}$ is solely through the material contribution in $\phi_{ind}$.  From the definition of the surface loss function \eqref{surface loss function}, only the imaginary contribution of the material response is needed and we can extract this quantity from the total vacuum potential $\phi$ as
\ba
\label{dielectric recipe}
g(q,\omega) &=& -\lb\frac{(1+\e)^2}{2A(q,\omega)}\rb\im\phi(q,\omega,z=0)
\\ &=& -\lb\frac{(1+\e)^2}{2A(q,\omega)}\rb\im \phi_{ind}(q,\omega,z=0)
\ea
even when the $\phi_{ind},\phi_{ext}$ aren't known separately.  Naturally, the (total) potential $\phi$ associated with the electric field can be obtained through a dielectric analysis of the semi-infinite Fetter model and the calculation of the surface loss function reduces to using  \eqref{dielectric recipe} on the vacuum solution.

\section{Results}
Upon applying the aforementioned electromagnetic boundary conditions to our implementation of the semi-infinite Fetter model, we can extract the surface loss function $g$ from the vacuum potential $\phi$ through the relation given in \eqref{dielectric recipe}.  The surface loss function can be separated into two components
\ba
\label{loss decomposition}
&\,& g(q,\omega) = g_s(q,\omega) + g_b(q,\omega)\,\,\,,
\ea
where
\ba
\label{surface loss}
&\,& g_s(q,\omega) := \frac{(1+\e)^2}{4}
\\ \nn &\,& \times\,
\im
\frac{2\bar q \e^2 s\cosh\bar q-\lp2\bar\omega^2\e^2+\bar q\e s\rp\sinh\bar q}{\bar q\e^2 s\cosh\bar q-\lb
\lp\e^2-1\rp\bar\omega^2+\bar q\e s
\rb\sinh\bar q}
\ea
and
\ba
\label{bulk loss} &\,& g_b(q,\omega) := \frac{(1+\e)^2}{4}
\\ \nn &\,&
\times\,\im
\frac{\e\sqrt{\sinh\bar q\lb
\lp
4\bar\omega^4+\bar q^2 s^2
\rp\sinh\bar q-4\bar q \bar\omega^2 s\cosh\bar q
\rb}}{\bar q\e^2 s\cosh\bar q-\lb
\lp\e^2-1\rp\bar\omega^2+\bar q\e s
\rb\sinh\bar q}
\ea
are loosely associated with surface ($g_s$) and bulk ($g_b$) plasmon excitation; similar decompositions have been previously noted \cite{Bergara1999,Pitarke2001,Jain1985_2}.  In the definition of $g_s$ \eqref{surface loss} and $g_b$ \eqref{bulk loss}, the dimensionless parameters $\bar q:=qa, \bar\omega:=\omega/\Omega, s:=(1+i/\omega\tau)^{-1}$ have been introduced, $\Omega$ is the optical plasma frequency \eqref{Omega def}, and the assumption $\e>1$ has been used.  Additionally, the complex square root in the definition of $g_b$ \eqref{bulk loss} corresponds to the branch with positive imaginary part.

\begin{figure*}
	\centering
	\subfloat[\label{sfig:gall_highq}]{
		\includegraphics[width=0.49\linewidth]{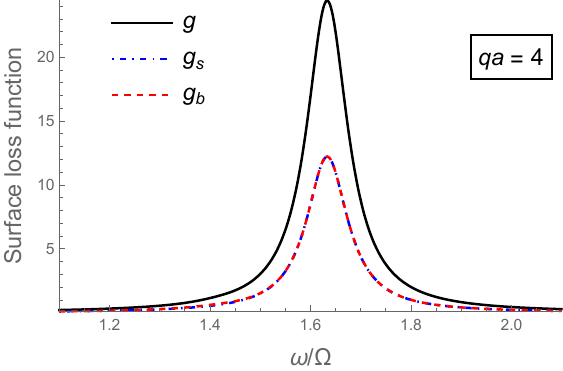}
	}
	\subfloat[\label{sfig:gall_medq}]{
		\includegraphics[width=0.49\linewidth]{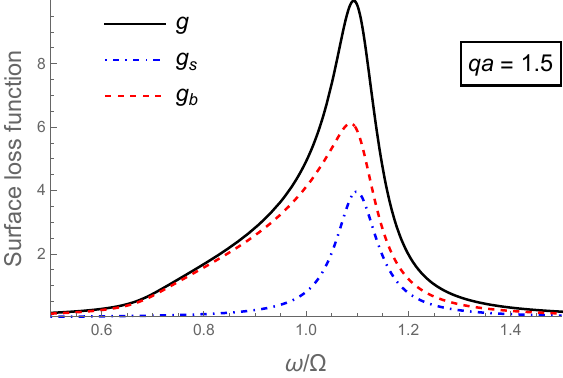}
	}
	
	\subfloat[\label{sfig:gall_medlowq}]{
		\includegraphics[width=0.49\linewidth]{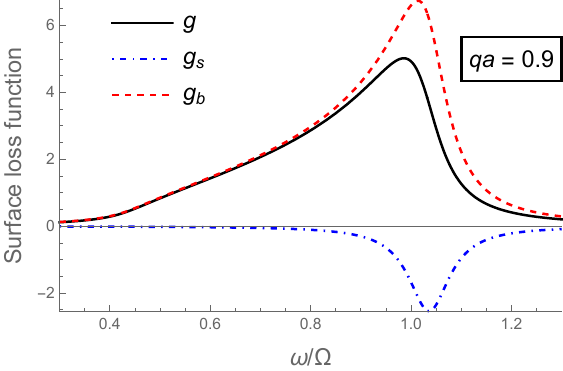}
	}
	\subfloat[\label{sfig:gall_lowq}]{
		\includegraphics[width=0.49\linewidth]{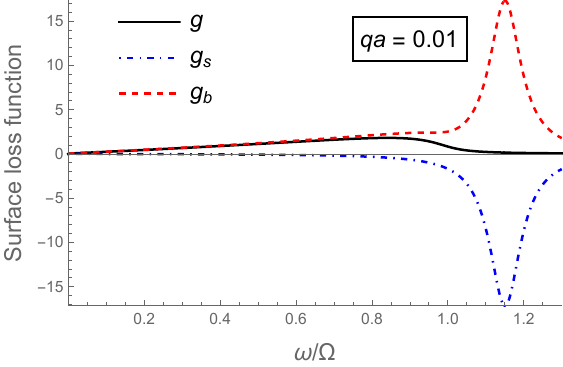}
	}
	
	\caption{Plotted above is the surface loss function, $g=g_s+g_b$ of \eqref{loss decomposition}, at $\e=2$ and $\tau=10/\Omega$ across several values of the in-plane wavevector $q$.  At $\e=2$, the cutoff wavevector is given by \eqref{cutoff q} as $q^*a=\ln 3\approx 1.1$.  At each wavevector the surface plasma frequency $\omega_{sp}$ and continuum of bulk plasma frequencies can be determined by \eqref{surface dispersion} and \eqref{bulk dispersion}, respectively.  In {\bf (a)}, $qa=4$ ($q>q^*$), $\omega_{sp}\approx1.63\,\Omega$, and the plasmon continuum spans $1.39\,\Omega<\omega<1.44\,\Omega$.  In {\bf (b)}, $qa=1.5$ ($q>q^*$), $\omega_{sp}\approx1.10\,\Omega$, and the plasmon continuum spans $0.69\,\Omega<\omega<1.09\,\Omega$.  In {\bf (c)}, $qa=0.9$ ($q<q^*$),  $\omega_{sp}\approx1.04\,\Omega$ is non-propagating, and the plasmon continuum spans $0.44\,\Omega<\omega<1.03\,\Omega$.  In {\bf (d)}, $qa=0.01$ ($q<q^*$) is the $qa\to0$ limit, $\omega_{sp}\approx1.15\,\Omega$ is non-propagating, and the plasmon continuum spans $0<\omega<\Omega$.}
	\label{figLoss}
\end{figure*}

At sufficiently large $q>q^*$, the essential behavior of the surface $g_s$ \eqref{surface loss} and bulk $g_b$ \eqref{bulk loss} contributions to the surface loss function $g$ can be gleaned from their identical pole structure.  Notably, both $g_s$ and $g_b$ have the same denominator that, in the $s\to1$ ($\tau\to\infty$) limit, can be written as
\ba
\nn &\,&
\frac{1}{\bar q\e^2\cosh\bar q-\lb
\lp\e^2-1\rp\bar\omega^2+\bar q\e
\rb\sinh\bar q}
\\
\label{denom}
&\,&
=
\lb \frac{1}{\lp \e^2-1\rp\sinh\bar q}\rb
\frac{1}{
\omega_{sp}^2(q)/\Omega^2-\bar\omega^2}
\ea
in terms of the surface plasma frequency, $\omega_{sp}$ of \eqref{surface dispersion}.  So long as the relaxation time $\tau$ is not short enough to radically alter the surface plasmon dispersion, {\it both} the surface $g_s$ and bulk $g_b$ terms peak at $\omega=\omega_{sp}$.  When the surface plasmon is a sharp resonance of the semi-infinite Fetter model (i.e., for $q$ sufficiently larger than $q^*$), both $g_s$ and $g_b$ appear similar in character and simply provide two contributions to the spectral weight at the surface plasma frequency $\omega_{sp}$.  This behavior can be seen at $qa=4$ in Fig. \ref{sfig:gall_highq}; the $g_s,g_b$ curves lie atop one another and the cutoff wavevector $q^*a=\ln 3\approx1.1$ (from \eqref{cutoff q} using $\e=2$ of Fig. \ref{figLoss}) is suitably smaller than $qa=4$.

When $q>q^*$ approaches $q^*$, the distinction between $g_s$ and $g_b$ becomes apparent as the square root in the numerator of $g_b$ \eqref{bulk loss} begins to appreciably contribute.  In the $s\to1$ ($\tau\to\infty$) limit, the quartic polynomial in $\bar\omega$ within the square root of \eqref{bulk loss},
\ba
\lp
4\bar\omega^4+\bar q^2
\rp\sinh\bar q -4\bar q \bar\omega^2\cosh\bar q\,\,\,,
\ea
has the zeroes
\ba
\label{zeroes}
\omega = \omega_p(q,Q=0)\,\,\, &\text{and}& \,\,\,\omega=\omega_p(q,Q=\pi/a)\,\,\,.
\ea
In \eqref{zeroes}, $\omega_p$ is the bulk plasma frequency given by \eqref{bulk dispersion}, and $Q$ labels the out-of-plane wavevector of the bulk plasmon in the {\it infinite} Fetter model.  Even in the $s\to1$ limit, the square root in the definition of $g_b$ \eqref{bulk loss} contributes a finite imaginary part when $\omega$ is within the bulk plasmon continuum, or $\omega_p(q,Q=\pi/a)<\omega<\omega_p(q,Q=0)$.  The bulk part ($g_b$) of the surface loss function $g$, then, is made up of {\it two} contributions: a peak at the surface plasma frequency $\omega=\omega_{sp}$ and a continuum of modes across the bulk plasmon dispersion $\omega_p(q,Q)$ \eqref{bulk dispersion} at fixed in-plane $q$.  The contribution of both bulk and surface modes can be observed in Fig. \ref{sfig:gall_medq} at $qa=1.5$ where the chosen value of $\e=2$ still maintains $qa>q^*a=\ln 3\approx 1.1$ from \eqref{cutoff q}.

For $q<q^*$, the surface plasmon is no longer a self-sustained resonance of the semi-infinite Fetter model: this is how $q^*$ in \eqref{cutoff q} is defined.  The lack of surface plasmon propagation is a precise statement occurring at $q^*$ so long as finite-$\tau$ corrections to the surface plasmon dispersion are negligible \cite{Quinn1983,Jain1985_1,Jain1985_2}; however, the spectrum itself evolves continuously across $q=q^*$ \cite{Jain1985_3}.  Nevertheless, there is a sharp, qualitative change in the surface contribution $g_s$ across $q=q^*$.  From the definition of $g_s$ \eqref{surface loss} at finite $\tau$ (i.e., {\it not} in the $s\to1$ limit), $g_s$ is proportional to the frequency-independent factor
\ba
\label{surface loss factor}
g_s(q,\omega) \propto\lb-2\e\cosh qa+\lp1+\e^2\rp\sinh qa\rb
\,\,\,,
\ea
which vanishes at the cutoff wavevector $q=q^*$ of \eqref{cutoff q}.  For $q>q^*$, the multiplicative factor in \eqref{surface loss factor} is positive whereas it is {\it negative} for $q<q^*$.  Naturally, the total surface loss function $g=g_s +g_b$ cannot become negative as the density response in \eqref{surface loss function} is related to a sum of transition amplitudes for charge density excitations via the fluctuation-dissipation theorem.  Instead, the sign change across $q=q^*$ represents a kind of inverse ``{\it begrenzungs} effect" \cite{Lucas1972,Egerton2011} wherein the surface component $g_s$ suppresses spectral weight at the (non-propagating) surface plasma frequency arising from the pole in $g_b$ at $\omega=\omega_{sp}$.  The net effect of the suppression at $\omega=\omega_{sp}$ is to increase the relative contribution from the bulk plasmon continuum to the (total) surface loss function $g$.  This behavior can be observed in Fig. \ref{sfig:gall_medlowq} at $qa=0.9<q^*a=\ln3\approx1.1$ for the chosen value $\e=2$.

\begin{figure}[h!]
	\centering
	\subfloat[\label{sfig:g_lowq}]{
		\includegraphics[width=0.49\linewidth]{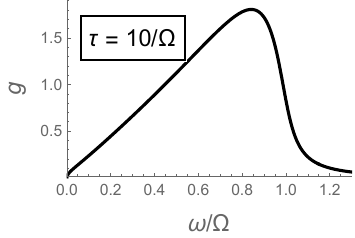}
	}
	\subfloat[\label{sfig:g_lowq_sharp}]{
		\includegraphics[width=0.49\linewidth]{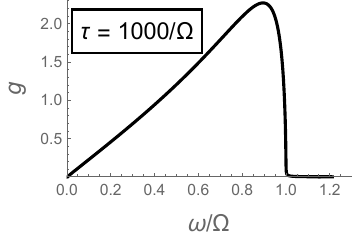}
	}
	
	\caption{Plotted above is the surface loss function $g=g_s+g_b$ at $\e=2$ and the $qa\to0$ limit implemented by $qa=0.01$.  In {\bf (a)}, $g$ from Fig. \ref{sfig:gall_lowq} ($\tau=10/\Omega$) is plotted on its own for clarity.  In {\bf (b)}, $g$ is plotted for the same $\e=2$ and $qa=0.01$, but at the much longer relaxation time $\tau=1000/\Omega$.  In the $qa\to0$ limit, the bulk plasmon continuum is bounded by $0<\omega<\Omega$ from \eqref{bulk dispersion}.}
	\label{figgLowq}
\end{figure}

\begin{figure}[h!]
	\centering
	\subfloat[\label{sfig:g_lowq_1p1}]{
		\includegraphics[width=0.49\linewidth]{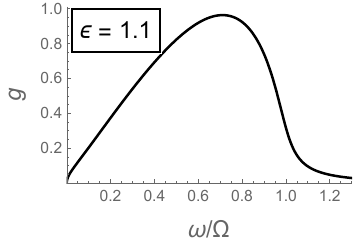}
	}
	\subfloat[\label{sfig:g_lowq_2}]{
		\includegraphics[width=0.49\linewidth]{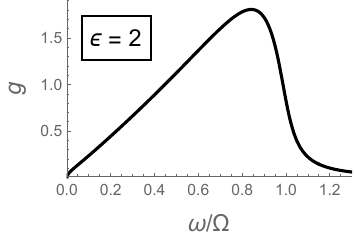}
	}
	
	\subfloat[\label{sfig:g_lowq_5}]{
		\includegraphics[width=0.49\linewidth]{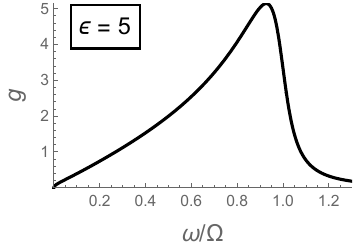}
	}
	\subfloat[\label{sfig:g_lowq_10}]{
		\includegraphics[width=0.49\linewidth]{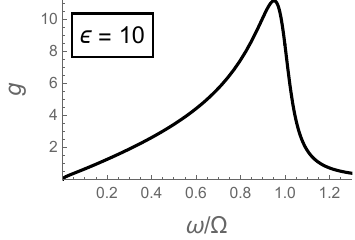}
	}
	\caption{Plotted above is the variation of the $qa\to0$ ($qa=0.01$) surface loss function $g$ \eqref{loss decomposition} with the dielectric constant $\e$.  The parameters used are $\tau=10/\Omega$ and: $\e=1.1$ ($q^*a\approx3.0$) in {\bf (a)}, $\e=2$ ($q^*a\approx1.1$) in {\bf (b)}, $\e=5$ ($q^*a\approx0.41$) in {\bf (c)}, $\e=10$ ($q^*a\approx 0.20$) in {\bf (d)}.  The value $\e$ determines $q^*$ through \eqref{cutoff q}.}
	\label{figgEps}
\end{figure}

In the $qa\to0$ limit, the non-propagating surface plasmon peak at $\omega=\omega_{sp}$ becomes increasingly suppressed, leaving only a broad response across the bulk plasmon continuum $0<\omega<\Omega$.  This behavior can be observed in Fig. \ref{sfig:gall_lowq} for $qa=0.01$ and $\e=2$: there is no appreciable feature in the surface loss function $g$ at the peak $\omega=\omega_{sp}$ visible for either $g_s$ or $g_b$.  As the vertical scale due to $g_b$ and $g_s$ obscures $g$ in Fig. \ref{sfig:gall_lowq}, $g$ is plotted by itself in Fig. \ref{sfig:g_lowq}.  Curiously, the broad response across the bulk plasmon continuum is relatively insensitive to the relaxation time $\tau$ so long as the plasmon dispersion is not noticeably altered from its free electron value in \eqref{bulk dispersion}.  In Fig. \ref{sfig:g_lowq_sharp}, the surface loss function $g$ is plotted for the same $qa\to0$ limit ($qa=0.01$) and dielectric constant $\e=2$, but at the long relaxation time $\tau=1000/\Omega$; notwithstanding, the broad shape across $0<\omega<\Omega$ remains and the only qualitative difference is the loss of curvature near $\omega=0,\,\Omega$ when compared to $g$ at $\tau=10/\Omega$ in Fig. \ref{sfig:g_lowq}.  The low energy, linear tail and broad peak near --- but not {\it at} --- the optical plasma frequency $\Omega$ appear as signatures of the underlying bulk plasmon continuum, rather than resulting from any particular damping factor.  While this statement is qualitatively general, it should be noted that the precise shape of the surface loss function in the $qa\to0$ limit is sensitive to the dielectric constant $\e$.  In Fig. \ref{figgEps}, the $qa\to0$ surface loss function is shown at $\e=1.1,2,5,10$ to demonstrate the influence of $\e$ on the spectrum.

\begin{figure}[h!]
	\centering
	\subfloat[\label{sfig:reeps}]{
		\includegraphics[width=0.49\linewidth]{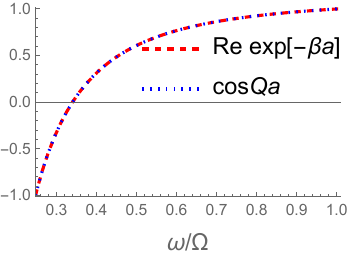}
	}
	\subfloat[\label{sfig:imeps}]{
		\includegraphics[width=0.49\linewidth]{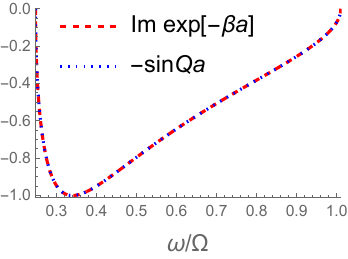}
	}
	
	\caption{Plotted above are the {\bf (a)} real and {\bf (b)} imaginary parts of the decay factor $\exp[-\beta a]$ to demonstrate the equality between the out-of-plane oscillation in the electric potential and the out-of-plane wavevector $Q$ \eqref{cos Q} across the plasmon continuum.  The parameters used are $qa=0.5$, $\e=2$, and the $\tau\to\infty$ limit.  In both {\bf (a)} and {\bf (b)}, the two curves lie atop one another.}
	\label{figDecay}
\end{figure}

In addition to the surface loss function having support across the (bulk) plasmon continuum, the induced potential at these energies oscillates in step with the associated plasmon.  Through our dielectric analysis, we can provide a precise, one-to-one correspondence between the plasmon at out-of-plane wavevector $Q$ --- $\omega_p(q,Q)$ of \eqref{bulk dispersion} --- and the material response at fixed $q$ across the plasmon continuum.  In the dielectric theory, this out-of-plane oscillation occurs in the decay factor $\exp[-\beta a]$, which relates the potential in adjacent dielectric layers: $\phi_n=\exp[-\beta a]\phi_{n+1}$.  In order to maintain a bounded solution as $n\to\infty$, we require $\re\beta\ge0\iff |\exp[-\beta a]|\le 1$; decaying modes (surface plasmons) correspond to $|\exp[-\beta a]|<1$ whereas wavelike modes (bulk plasmons) correspond to $|\exp[-\beta a]|=1$ \cite{Quinn1983,Cottam1993,Cottam2004}.  In the $\tau\to\infty$ limit of negligible conductivity dissipation, the decay factor across the plasmon continuum is determined solely by the out-of-plane oscillation of the (bulk) plasmon.  Specifically, rearranging the plasmon dispersion $\omega_p$ in \eqref{bulk dispersion},
\ba
\label{cos Q}
\cos Qa =
\frac{1}{2}\lb
2\cosh qa -\lp\frac{\Omega}{\omega}\rp^2qa\sinh qa
\rb\,\,\,,
\ea
enables us to label the bulk plasmon at fixed $(q,\omega)$ through its out-of-plane wavevector $Q$.  Using the prescription in \eqref{cos Q}, we find that $\exp[-\beta a]=\exp[-i Qa]$ at energies satisfying the plasmon dispersion $\omega_p(q,Q)$ of \eqref{bulk dispersion}.  In Fig. \ref{figDecay}, we demonstrate this behavior by plotting the decay factor $\exp[-\beta a]$ across the bulk plasmon continuum at $qa=0.5$, $\e=2$, and in the $\tau\to\infty$ limit.  For clarity, the equivalence $\exp[-\beta a]=\exp[-i Qa]$ is maintained at arbitrary in-plane wavevector $qa$ and dielectric constant $\e>1$ across the bulk plasmon continuum.  The surface loss function, then, is a direct probe of the (bulk) plasmon across $\omega_p(q,Q)$ at fixed in-plane $q$.

The surface loss function $g$ \eqref{loss decomposition} demonstrates the transition from surface plasmon propagation at large in-plane wavevectors to a long wavelength regime dominated by the bulk plasmon continuum (see Fig. \ref{figLoss}).  Instead of a sharp peak at a single frequency, the $qa\to0$ surface loss function is characterized by a linear tail at low energies and a broad peak across the (bulk) plasmon continuum.  This shape is only weakly influenced by the intrinsic damping parameter (see Fig. \ref{figgLowq}); however, the spectrum is sensitive to the dielectric constant $\e$ (see Fig. \ref{figgEps}).  Though the bulbous peak might initially suggest a damped, residual surface plasmon at long wavelengths, the oscillatory behavior of the electric potential across the plasmon continuum (e.g., the exact correspondence $\exp[-\beta a]=\exp[-i Qa]$ demonstrated in Fig. \ref{figDecay}) suggests that the broad spectrum is intrinsic to the plasmon itself.  From this perspective, we find that the long wavelength surface loss function is purely a {\it bulk} probe of the semi-infinite Fetter model.

\section{Discussion}
Previous studies \cite{Bergara1999,Pitarke2001,Mills1984,Mills1994,Lucas1985,Lucas1985_2,Lucas1985_3,Lambin1987,Jain1985_2,Jain1985_3} have demonstrated, and emphasized, that bulk charge density excitations appear in the surface loss function (i.e., the HREELS cross section); however, the contribution of bulk excitations typically presents as either a weak, possibly imperceptible, shoulder near sharp surface excitations or a broad background.  We found similar behavior in the semi-infinite Fetter model for sufficiently large (in-plane) wavevectors.  At $qa=1.5>q^*a$ in Fig. \ref{sfig:gall_medq}, a shoulder forms across the bulk plasmon continuum before the strong peak at the surface plasma frequency.  Just below the cutoff wavevector ($q<q^*$) in Fig. \ref{sfig:gall_medlowq}, however, the peak at the (non-propagating) surface plasma frequency has become suppressed and the surface loss function takes on a broad shape across the bulk plasmon continuum.  In the long wavelength limit ($qa\to0$), the peak at the (non-propagating) surface plasma frequency is visible in both $g_s$ and $g_b$ in Fig. \ref{sfig:gall_lowq}; yet, no such feature appears in the surface loss function, $g=g_s+g_b$, of Fig. \ref{sfig:g_lowq}.  In essence, the long wavelength surface loss function of the semi-infinite Fetter model probes {\it only} the bulk plasmon continuum of the infinite system.

In the absence of a propagating surface plasmon, the transfer of spectral weight to the bulk plasmon continuum is {\it required} by the f-sum rule.  While we have performed a dielectric analysis, the electromagnetic response encoded within the vacuum potential \eqref{dielectric recipe} genuinely behaves as the many-body response function defined in \eqref{surface loss function}.  In particular, the f-sum rule,
\ba
\label{f sum def}
\int_0^\infty d\omega\,\omega\, g(q,\omega) = \frac{\pi e^2|q|}{4\e_0m}\int_0^\infty dz\, n(z) e^{-2|q|z}
\,\,\,,
\ea
relates the spectral weight within the surface loss function $g$ \eqref{surface loss function} of our layered model to the electron density $n(z)$.  Notably, the f-sum rule in \eqref{f sum def} is reduced when compared to its standard form \cite{Pitarke2001,Liebsch1997,Gumhalter1984,Gumhalter1975} due to the lack of out-of-plane dispersion in the semi-infinite Fetter model (e.g., see the derivation in \cite{Gumhalter1976}).  As a system of regularly-spaced conducting planes, the electron density in our analysis is simply $n(z)=n\sum_j \delta(z-ja)$, where $n$ is the planar electron density of a single layer, and the right-hand side of the f-sum rule in \eqref{f sum def} can be evaluated as a geometric series to provide
\ba
\label{f sum}
\int_0^\infty d\omega\,\omega\, g(q,\omega) = \lp\frac{\pi\Omega^2_0}{8}\rp\frac{2|q|a}{1-e^{-2|q|a}}\,\,\,,
\ea
where
\ba
\Omega^2_0:=\frac{ne^2}{\e_0 ma}=\e\,\Omega^2
\ea
is the unscreened optical plasma frequency.

\begin{figure}[h!]
	\centering
	\includegraphics[width=\linewidth]{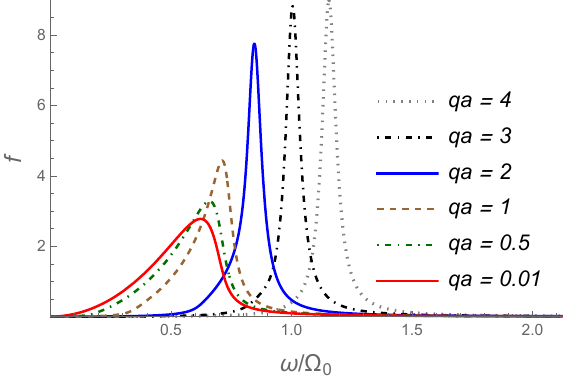}
	\caption{Plotted above is the redistribution of spectral weight from the surface plasma frequency to the (bulk) plasmon continuum in the scaled surface loss function $f$ \eqref{f def}.  The parameters used are $\e=2$, $\tau=10/\Omega$, and the cutoff wavevector (determined through \eqref{cutoff q} by $\e=2$) $q^*a=\ln3\approx1.1$.  By construction, all curves have the same (unit) integrated weight.  The frequency axis has been scaled by $\Omega_0=\sqrt\e\,\Omega$ to demonstrate the normalization; in these units, the plasmon continuum spans $0<\omega<\Omega_0/\sqrt\e$ in the long wavelength limit.}
	\label{figfsum}
\end{figure}

The crucial consequence of the f-sum rule \eqref{f sum} is that the $qa\to0$ limit of the surface loss function remains finite:
\ba
\label{f sum isotropic}
\lim_{qa\to0}\int_0^\infty d\omega\,\omega\, g(q,\omega) = \lp\frac{\pi\Omega^2_0}{8}\rp
\,\,\,.
\ea
The sharp $q\to0$ surface plasmon in an isotropic metal exhausts the spectral weight required by the f-sum rule \cite{Pitarke2001,Gumhalter1975}; however, the semi-infinite Fetter model lacks this mode in the long wavelength limit.  Nevertheless, the f-sum rule is indifferent to whether or not sharp excitations are present: the contribution from the bulk plasmon continuum {\it must} compensate for the lost spectral weight at the surface plasmon peak.  To demonstrate how this spectral weight is redistributed across the cutoff wavevector $q^*$, we can scale the surface loss function according to the f-sum rule \eqref{f sum} as
\ba
\label{f def}
f(q,\omega):=\lb\lp\frac{8}{\pi\Omega_0^2}\rp\frac{1-e^{-2|q|a}}{2|q|a}\rb \omega\,g(q,\omega)
\ea
so that $f$ has unit weight: $\int_0^\infty d\omega\,f(q,\omega) =1$.  One should note that the re-scaling in \eqref{f def} slightly distorts the spectrum of $f$ (compared to the surface loss function $g$) due to the multiplicative factor of $\omega$.  In Fig. \ref{figfsum}, the dispersion of $f$ at $\e=2$ and $\tau=10/\Omega$ is shown across several values of $qa$ to demonstrate the rearrangement of spectral weight as the surface plasmon peak is suppressed below $q^*a=\ln 3\approx1.1$ (using \eqref{cutoff q} for $\e=2$).

Before comparing our results with existing data, we should acknowledge that the model presented in section \ref{section 1} --- the semi-infinite Fetter model (see Fig. \ref{figLattice}) --- contains several simplifying approximations.  When describing the conducting planes, the planar conductivity $\sigma(\omega)$ \eqref{Drude conductivity} was characterized by a constant relaxation time $\tau$ up to --- or, in the $q>q^*$ regime, {\it beyond} --- the plasma frequency and we neglected dispersion with the planar wavevector $q$.  The lack of dispersion means that we should restrict our analysis to the long wavelength limit, $ql\ll 1$, where $l$ is the planar lattice spacing.  If the inter-layer distance between conducting planes, $a$, is sufficiently larger than the planar lattice spacing (i.e., $l\ll a$), then the surface loss function, $g$ of \eqref{loss decomposition}, can disperse with $qa$ while maintaining $ql\ll1$.  Additionally, we chose to model the insulating layers by frequency-independent dielectric constants.  So long as the band gap of the dielectric layers is sufficiently large compared to the bulk or surface plasmon energies under consideration, we are primarily neglecting the low-energy phonon contributions of these insulating regions.

The approximations used in our simplified Fetter model of section \ref{section 1} are consistent with a long wavelength analysis of the layered structure in the high-$T_C$ cuprate superconductors.  To be specific, we will consider the structure of Bi-2212 (also modeled within a system of conducting layers in \cite{Mills1994}), which is characterized by the inter-bi-layer distance between copper oxide planes $a\approx 15.4$ \AA$\,$ and the planar (nearest-neighbor Cu-Cu) lattice spacing $l\approx3.8$ \AA$\,$ \cite{Springer2016}).  The small value $l/a\approx0.25$ permits the surface loss function, $g$ of \eqref{loss decomposition}, to disperse with $qa$ while remaining in the (planar) long wavelength limit of $ql\ll1$.  While the low-frequency scattering rate, $1/\tau(\omega)$, is famously linear in cuprates near optimal doping, the linearity of $1/\tau(\omega)$ in Bi-2212 begins to deviate near excitation energies of $\hbar\omega\sim0.37$ eV \cite{vanderMarel2003}.  The relaxation time, $\tau(\omega)$, eventually saturates to a constant near $\hbar\omega\sim0.62$ eV  \cite{vanderMarel2003}, which is well below the optical plasmon energy of doped Bi-2212 ($\hbar\Omega\sim 1$ eV from the optically-determined loss function \cite{Bozovic1990,vanderMarel2016,Schulte2002}).  More subtly, Bi-2212 is a bi-layer cuprate with {\it two} closely-spaced copper oxide planes repeated every $a\approx15.4$ \AA$\,$ (i.e., we neglect the planar lattice shift of the full $c\approx 30.8$\AA$\,$ unit cell \cite{Springer2016}).  This bi-layer structure in a layered electron gas results in an additional, narrow band of acoustic modes \cite{Cottam2004,Quinn1983_2,Schulte1999} below the (bulk) plasmon continuum; however, the dispersion shown in Fig. \ref{bulk dispersion} remains intact and qualitatively unaltered.  As a result, we will neglect these lower-energy acoustic modes and focus on the broader plasmon continuum that occurs in the single-layer Fetter model of Fig. \ref{figLattice}.

HREELS measurements of Bi-2212 at long wavelengths \cite{Schulte2002,Mitrano2018,Ali2020} corroborate the broad spectrum and (low energy) linear tail of our dielectric model, which can be seen in Figs. \ref{figgLowq} and \ref{figgEps} at various relaxation times and dielectric constants, respectively.  Due to the large inter-bi-layer spacing of Bi-2212, the long wavelength regime of the surface loss function is set by the small wavevector $1/a\approx0.06$ \AA$^{-1}$; using \eqref{cutoff q}, we can approximate the cutoff wavevector as $q^*\approx0.03$ \AA$^{-1}$ from the value $\e\approx4.5$ \cite{Schulte2002,vanderMarel2003,vanderMarel2016}.  In this regime, HREELS studies \cite{Schulte2002,Mitrano2018,Ali2020} observe a broad peak --- i.e., a peak with FWHM wider than the plasmon peak in Bi-2212 inferred via optical means \cite{Schulte2002,Presura2003,Bozovic1991,vanderMarel2016} --- centered near the optical plasma frequency which, in our model, is associated with the broad spectrum of {\it bulk} plasmon excitation (i.e., the dispersion in Fig. \ref{figBulk} at small $qa$).  Unfortunately, significant damping --- either the result of the (insulating) cap layer \cite{Mills1994} or some otherwise novel cuprate behavior --- at larger wavevectors (detailed in \cite{Mitrano2018,Ali2020}) prevents tracking the dispersion of this broad spectrum beyond the long wavelength limit; consequently, the regime of surface plasmon propagation for $q\gg q^*$ is not observed.  This discrepancy suggests that the planar conductivity --- or, as discussed in \cite{Mitrano2018,Ali2020}, the polarizability --- disperses in a nontrivial manner beyond the long wavelength limit.  As an additional complication, the low energy incident electrons used in HREELS result in a skewed spectrum for long wavelength scattering near the wavevector resolution  \cite{Schulte2002,Palmer1987,Zhu2022}.  The peak structure of broad, eV-scale losses can be entirely washed out at these long wavelengths \cite{Schulte2002,Ali2020}, rendering the surface loss function difficult to extract in Bi-2212.  Nevertheless, correcting for this distorted spectrum appears to recover a long wavelength surface loss function that scales \textit{linearly} at low energies \cite{Schulte2002}, rather than the quadratic tail observed in optical studies of the (bulk) loss function \cite{Bozovic1990,vanderMarel2016}.  In our model, this linear tail is associated with the broad plasmon continuum at long wavelengths (see Fig. \ref{figBulk}).  Further, both a broad peak near the optical plasma frequency and (low energy) linear tail were observed in the HREELS spectrum of the related {\it single}-layer compound Bi-2201 \cite{Schulte2002}, which suggests that these features are related to the cuprate layering structure.

\section*{Summary}
Electron Energy Loss Spectroscopy (EELS) in a reflection geometry (High Resolution EELS or HREELS) is often understood as a surface probe whose characteristic long wavelength excitation is the surface plasmon.  In the semi-infinite Fetter model (see Fig. \ref{figLattice}), however, the extreme limit of conduction anisotropy restricts surface plasmon propagation to finite in-plane wavevectors $q$ above a cutoff value $q^*$.  Below $q^*$, the system lacks a surface plasmon and, therefore, has only the (bulk) plasmon as its collective, long wavelength excitation.  As the plane of reflection explicitly breaks translation symmetry, the plasmon contributes a {\it continuum} of excitations due to the non-conserved out-of-plane wavevector.  To understand how this continuum might be observed in an HREELS experiment, we have employed a dielectric analysis to calculate the surface loss function --- the material response probed by HREELS --- of the semi-infinite Fetter model.

Our results can largely be summarized by Fig. \ref{figLoss}.  At sufficiently large $q>q^*$ in Figs. \ref{sfig:gall_highq} and \ref{sfig:gall_medq}, the surface loss function $g$ of \eqref{loss decomposition} is characterized by a sharp surface plasmon peak with only a weak (or, in Fig. \ref{sfig:gall_highq}, imperceptible) shoulder across the plasmon continuum.  Just below $q^*$ in Fig. \ref{sfig:gall_medlowq}, the surface plasmon peak becomes actively suppressed and the plasmon continuum receives significant spectral weight.  In Fig. \ref{sfig:gall_lowq}, the long wavelength limit has successfully suppressed the non-propagating surface plasmon peak and all that remains in the surface loss function is a broad response across the plasmon continuum (see Fig. \ref{sfig:g_lowq} or \ref{sfig:g_lowq_2}).  While the particular shape of the surface loss function requires a calculation, our results can be understood within the context of the f-sum rule \eqref{f sum def} governing the semi-infinite Fetter model \eqref{f sum}.  Whether or not there exist sharp excitations in the long wavelength limit, the surface loss function must contain significant spectral weight \eqref{f sum isotropic}.  Consequently, the long wavelength surface loss function becomes a probe of the {\it bulk} plasmon continuum in the semi-infinite Fetter model.

\section*{Acknowledgments}
The authors would like to thank Ali Husain, Karina Th\aa nell, Yu He, Edwin Huang, Xuetao Zhu, Abhishek Nag, Ke-Jin Zhou, Jin Chen, and Peter Abbamonte for helpful discussions.  CB and LY were funded partially by DMR-1919143 and PWP was funded partially by DMR-2111379.

\Urlmuskip=0mu plus 1mu
\bibliographystyle{apsrev4-2}
\bibliography{mybib}

\end{document}